\documentclass[amsmath,twocolumn,aps,prb]{revtex4}
\usepackage{bm}
\usepackage{color}
\usepackage{graphicx}
\newcommand{\bq}{\begin{equation}}
\newcommand{\ee}{\end{equation}}
\begin{document}

\title{Long-range exchange interaction between magnetic impurities in graphene}

\author{M. Agarwal}
\affiliation{Department of Physics and Astronomy, University of Utah, Salt Lake City, UT 84112, USA}

\author{E. G. Mishchenko}
\affiliation{Department of Physics and Astronomy, University of Utah, Salt Lake City, UT 84112, USA}

\begin{abstract}
The effective spin exchange coupling between impurities (adatoms) on graphene mediated by conduction electrons  is studied as a function of the strength of the  potential part of the on-site energy  $U$ of the electron-adatom interaction. With increasing $U$, the exchange coupling becomes long-range, determined largely by the impurity levels with energies close to the  Dirac points. When adatoms reside on opposite sublattices, their exchange coupling, normally antiferromagnetic,  becomes ferromagnetic and  resonantly enhanced at a specific distance where an impurity level crosses the Dirac point.
\end{abstract}

 \maketitle

\section{Introduction}

Among possible technological promises of graphene\cite{CN} are
both electronic and magnetic applications. The former include
transistors and  require control of graphene's conduction,
while the latter aim to build memory devices and hinge on the
ability to create and manipulate local magnetic moments.
Nonetheless, both these avenues actively explore the
possibilities of controlling graphene properties with chemical
dopants, such as hydrogen.

In electronic applications the role of dopants is to  suppress
otherwise strong metallic conductivity of the
material\cite{BME,ENM,BJN,DQF}. Magnetic
impurities\cite{LFM,PGS,YH,Yaz,BDS,SAS, EWG,EWT,GEW} typically
interact with the conduction band and produce electronic states
that carry magnetic moments and could be spread\cite{GGM} over
many lattice spacings $a$. Of particular interest is the mutual
interaction between such impurities. The potential part of this
interaction is important in determining the equilibrium spatial
arrangement of the dopants\cite{CSA,ASL,KCA}. At the same time,
the type of collective magnetic properties is sensitive to the
effective exchange coupling between dopants. Both the potential
part of the effective interaction energy and its spin-dependent
part are mediated by the conduction $\pi$-electrons of
graphene.

\begin{figure}
        \centering
    	\includegraphics[scale = 0.50]{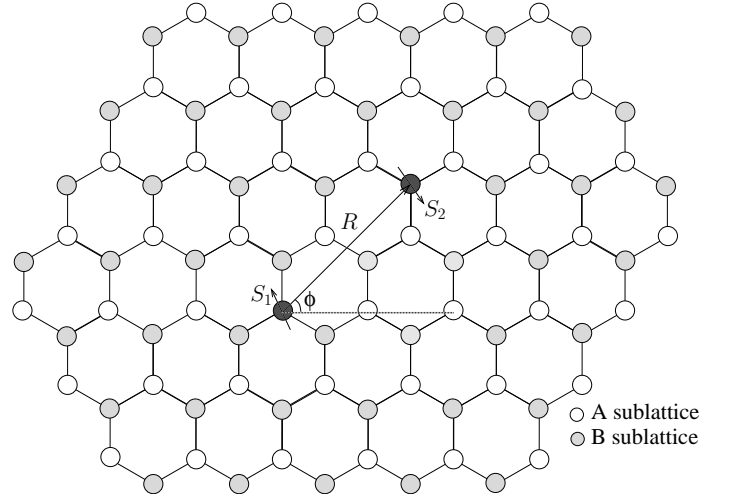}
	\caption{Graphene structure consisting of two sublattices A and B. Two magnetic impurities (shown in dark grey), with spins ${\bf S}_{1}$ and ${\bf S}_{2}$ are separated by the vector ${\bf R}$. The angle $\phi$ is counted from a zigzag direction.}
	\label{fig 1}
\end{figure}

One particularly promising dopant is hydrogen. Because it has
an energy level close to the Dirac point of the conduction
$\pi$-band of graphene \cite{WKL} coupling of conduction
electrons has a resonant character, whose scattering amplitude
resembles that of a strong substitution impurity\cite{MM}.
Motivated by this similarity, we are going to concentrate on
the substitution model, where the interaction of the impurity
with conduction electrons,
\begin{equation}
\label{energy of one impurity}
\hat H_{\rm imp}= U + J {\bf S} \cdot \hat {\bm \sigma },
\end{equation}
has both the  on-site potential energy $U$ and the spin part
described by the exchange coupling constant $J$; the spin
operator of the impurity ${\bf S}$ couples to the local value
of the conduction electron spin density $\hat {\bm \sigma}$.

When two dopants reside above carbon lattice sites separated by
the radius-vector ${\bf R}$, their effective
electron-mediated interaction, likewise, has two parts,
\begin{equation}
\label{interaction of two impurities definition}
\hat H_{12} =  W({\bf R}) +J_{\rm eff} ({\bf R}) {\bf S}_1 \cdot {\bf S}_2,
\end{equation}
The potential part $W({\bf R})$ has been studied both for
weak\cite{LTM} and strong\cite{SAL,LTM} impurity strength $U$.
To the contrary, the indirect spin exchange $J_{\rm eff}({\bf R})$
in graphene, while extensively studied perturbatively within
the usual RKKY approach\cite{Sar,BFS,BS,GKF,SS,Kog,PF}, has not
been addressed for impurities with large $U$. Two notable
previous research directions should be mentioned in this regard. In
Ref.~\onlinecite{KRA} the indirect exchange  between resonant
Anderson impurities in graphene was studied numerically with
the emphasis on the short-distance behavior. No analytic
dependence has been reported, however, in the strong coupling
limit. In Refs.~\onlinecite{BB,CZW} the indirect spin exchange
in the strong $U$ limit has been addressed in a somewhat
reminiscent situation of a topological insulator, but two
related features distinguish graphene from that case.

First, in the low-energy Hamiltonian of a topological
insulator, $H_0=v\hat {\bm \sigma} \cdot {\bf p}$, the spin
operator $\hat {\bm \sigma}$ is essentially the same as (or at
least directly related to)  the corresponding operator in
Eq.~(\ref{energy of one impurity}). In graphene, in contrast,
the Pauli matrices in the Hamiltonian relate to the pseudospin
operator, acting in the sublattice space. Second, in graphene the
interference of the electron states from the two Dirac points
results in {\it different} signs of the inter-dopant
interaction when the two dopants resite on the {\it opposite}
sublattices (AB-case) as compared with the {\it same }
sublattice (AA) arrangement. This complication did not arise
in Refs.~\onlinecite{BB,CZW}. In the present paper we
are going to investigate the indirect spin exchange coupling
$J_{\rm eff}({\bf R})$ between two dopants residing directly above
carbon atoms of intrinsic graphene. We study how $J_{\rm eff}({\bf R})$ depends on
1) the sublattice arrangement of the dopants, and 2)  the
strength of the potential coupling $U$, not assumed to be
small. At the same time, it will be sufficient to limit our
analysis to the lowest (second) order in the coupling constant
$J$, which is presumably never too large.

Let us first recapitulate the existing results for intrinsic
graphene with the Fermi level at the Dirac points. To the
lowest second order in $U$ and $J$, the results are rather
identical (barring the replacement $U^2 \leftrightarrow J^2$)
for both $W({\bf R})$ and $J_{\rm eff}({\bf R})$. When the two
dopants reside on the same sublattice, both quantities are
$\propto |{\bf R}|^{-3}$ and are {\it negative}, meaning
that the dopants {\it attract} and favor {\it ferromagnetic}
arrangement of their spins. On the other hand, the sign of both
interactions is reversed in the AB-case: $W({\bf R})>0$
and $J_{\rm eff}({\bf R})>0$, indicating {\it repulsion} and {\it
anti-ferromagnetic} coupling. The dependence on the distance
remains the same but the overall magnitude is three times
stronger than in the AA-case.

When the strength $U$ of the impurity becomes large, $U \gg
\hbar v |{\bf R}|/a^2$, the potential part $W({\bf
R})$ changes sign, compared with the weak coupling limit.
This happens for both AA and AB arrangements. This sign
reversal is the feature of the linear Dirac spectrum.
Additionally, the dependance of the inter-dopant interaction
becomes long-range\cite{SAL,LTM} $W({\bf R})\propto |{\bf
R}|^{-1}$ (up to additional logarithmic factors). It turns
out that the interaction energy in the AB-case is no longer
 stronger by a mere numerical factor, compared with the
AA configuration  (as is the case for weak $U$), but exceeds it
logarithmically, by a large factor $\ln{(|{\bf R}|/a})$.

Our goal is to perform a corresponding analysis of  $J_{\rm eff}({\bf
R})$. Below we carry it out  for both sublattice
configurations and for any strength of the potential $U$
assuming only  that $|{\bf R}| \gg a$. In Section II the energy spectrum in the presence of two impurities is discussed.   In Section III the
general expressions for the effective exchange constant are derived in terms of the integrals over
the energy of  conduction electrons. In Section IV those
general expressions are evaluated in the strong coupling limit.

\section{Energy levels of a two-impurity system}

To understand how and why the effective interaction of the dopants is sensitive to their relative positions on the two sublattices, it is necessary to discuss the differences in  the electronic spectra induced by the presence of two dopants with a strong on-site potential coupling $U$, elucidated in Ref.~\onlinecite{LTM}. When only one potential impurity is present,  it may induce a low-energy state, provided that $U$ is strong enough. The energy of the state is determined by the equation,
\begin{equation}\label{energy of a single impurity}
  1+\frac{UA_0E}{\pi v^2}\Bigl[\ln{(t/|E|)} +i\pi/2 \Bigr] =0,
\end{equation}
where $A_0=3\sqrt{3}a^2/2$ is the  area of a graphene unit cell; and the bandwidth of the conduction band $t \sim v/a$. The energy $E \ll t$ if the on-site energy is large: $U \gg t$. The energy level is quasi-localized -- the overlap with the conduction band causes it to have a finite lifetime,
\begin{equation}\label{energy of a single impurity value}
E= -\frac{\pi v^2}{U\!A_0}\frac{1}{\ln{(|U|/t)} +i\pi/2}.
\end{equation}
Note that the impurity level is located in the lower Dirac cone, $E<0$, for repulsive potentials $U>0$, and vice versa.

When two impurities are present, their energy levels split. For the impurities located on the same sublattice (AA), the two energies are given by the equation,
\begin{align}\label{energy of two impurities AA}
  1+\frac{UA_0E}{\pi v^2}\Bigl[&\ln{(t/|E|)} +i\pi/2 \Bigr] \nonumber\\ &=\pm \frac{UA_0E}{\pi v^2}|\cos\theta_{AA}| \ln{\left(\frac{v}{R|E|} \right)},
\end{align}
where the parameter $ \theta_{\! AA} ({\bf R}) = \frac{2\pi
R}{3\! \sqrt{3}a} \cos\phi$ is equal to the phase difference
that the states belonging to different Dirac cones acquire when
they travel between the two impurities. It depends on both the
length of the radius-vector ${\bf R}$ and the angle $\phi$ it
makes with the zigzag direction, see Fig.~\ref{fig 1}. The oscillations in the
right-hand side described by this phase,
therefore, are caused by the interference of the wave functions
of different Dirac species.

Importantly, both solutions of Eq.~(\ref{energy of two impurities AA}) remain on the same side of the Dirac point $E=0$ as the single-impurity level (\ref{energy of a single impurity value}): no level can ``escape'' from its Dirac cone. This is most simply seen from the fact that no solution of Eq.~(\ref{energy of two impurities AA}) can ever have zero energy, $E=0$.

To the contrary, when two impurities reside on different sublattices (AB configuration), the energy levels are determined from the equation,
\begin{align}\label{energy of two impurities AB}
  1+\frac{UA_0E}{\pi v^2}\Bigl[\ln{(t/|E|)} +i\pi/2 \Bigr] =\pm \frac{UA_0}{\pi v R}|\sin\theta_{AB}|,
\end{align}
with the new phase being ${\theta_{\! AB}}({\bf R})=\phi+\frac{2\pi
R}{3\! \sqrt{3}a} \cos\phi$. The main new feature of the AB case is the existence of the $E=0$ solution at a specific distances $R$. Such solution, from Eq.~(\ref{energy of two impurities AB}), has the energy (for positive $U>0$),
\begin{equation}\label{energy of a crossing level}
  E= \frac{v(R_0-R)|\sin\theta_{AB}|}{RR_0(\ln{\frac{RR_0}{a|R_0-R|} +\frac{i\pi}{2})}}, \hspace{0.5cm} R_0 = \frac{U A_0}{\pi v}|\sin\theta_{AB}|,
\end{equation}
and passes from the lower Dirac cone to the upper cone as the distance between the adatoms becomes shorter than a {\it resonant distance} $R_0$. Right at $R=R_0$ one of the impurity levels lies exactly at the energy $E=0$. Additionally, the width of this level becomes vanishingly small. As we are going to see below, the effective exchange coupling between adatoms spins is resonantly enhanced when they are separated by the distance $R_0$.

\section{General expression for the indirect exchange coupling}

The Hamiltonian of  two impurities (adatoms) on graphene, one
positioned above a carbon atom at the origin (the sublattice it
belongs to is called $A$) and the second above the atom
some distance ${\bf R}$ away (which could belong to either of
the two sublattices), consists of three parts, $H = H_0 +H' + H'_{\rm sp}$, namely,
\begin{align}
\label{Hamiltonian}
&H_0 = t\sum_{{\bf r}}\sum_{i=1,2,3} \hat \psi^{\dagger}({\bf
r})\hat \psi({\bf r}+{\bf
a}_i),\nonumber\\
& H' = U\hat \psi^{\dagger} ({\bf 0})\hat \psi({\bf 0})+ U\hat \psi^{\dagger}({\bf R})\hat \psi({\bf R}), \nonumber\\
& H'_{\rm sp} =   J {\bf S}_1 \cdot \hat\psi^{\dagger}({\bf
0})\hat {\bm \sigma} \hat \psi({\bf 0}) + J {\bf S}_2 \cdot
\hat\psi^{\dagger}({\bf R})\hat {\bm \sigma} \hat \psi({\bf R}),
\end{align}
Here $H_0$ is the kinetic energy of electrons, the hopping integral $t$ being
assumed real and positive; {\it U} is the additional on-site
potential energy induced by an impurity; {\it J} is the
exchange coupling of an impurity spin  with the spin of the
conduction electrons. The hat above the electron operator $\hat
\psi$ indicates a spinor, the summation over the spin indices
is implied;  $\hat {\bm \sigma}$ are the Pauli matrices acting in the  spin (not pseudospin) space.

Because the spin exchange is never very strong, it is sufficient
to determine the effective impurity-impurity spin coupling
$J_{\rm eff}$ to the lowest (second) order in $J$. Most simply
this can be done by using the standard quantum-mechanical theorem that poses that the derivative of the impurity-impurity interaction energy (\ref{interaction of two impurities definition}) with respect to the coupling constant $J$ is equal to the expectation value of the corresponding derivative of the Hamiltonian
(\ref{Hamiltonian}):
\begin{multline}
\label{first_identity} \frac{\partial J_{\rm eff}}{\partial J}
{\bf S}_1 \cdot {\bf S}_2 = \left\langle \frac{\partial
H}{\partial J}\right\rangle= {\bf S}_1 \cdot\left\langle
\hat\psi^{\dagger}({\bf 0})\hat {\bm \sigma} \hat \psi({\bf 0})
\right\rangle \\ +{\bf S}_2 \cdot \left\langle
\hat\psi^{\dagger}({\bf R})\hat {\bm \sigma} \hat \psi({\bf
R})\right\rangle =
-i {\bf S}_1({\bf 0})\cdot {\rm Tr} \,\hat {\bm \sigma}\hat {\cal G} (0,0,t=-0) \\
-    i{\bf S}_2({\bf R})\cdot {\rm Tr} \,\hat {\bm
\sigma}\hat{\cal G} ({\bf R},{\bf R},t=-0),
\end{multline}
where we introduced Green's function of the system, which in
the interaction representation with respect to the spin part of
the Hamiltonian is (spin indices shown explicitly)
\begin{equation}
\label{greensfunction} {\cal G_{\alpha\beta}}({\bf r},{\bf r'},t) =-i\langle T \psi_{\alpha}({\bf r},t)
 \psi_{\beta}^\dagger({\bf r'},0)S(\infty, -\infty) \rangle,
\end{equation}
with the interaction matrix given by the standard expression,
\begin{equation}
\label{Smatrix}
S(\infty, -\infty)= T \exp{\left(\frac{-i}{\hbar}\int_{-\infty}^{\infty} H'_{\rm sp} dt\right)}.
\end{equation}
Expanding the $S$-matrix to the first order in the spin
Hamiltonian, we obtain,
\begin{equation}
\label{partial_Jeff}
\frac{\partial J_{\rm eff}}{\partial J}= -4iJ\int\limits_{-\infty}^{\infty} \frac{dE}{2\pi \hbar}
                                   {\cal G}_E({\bf R},0){\cal G}_E(0,{\bf R}).
\end{equation}
Here, ${\cal G}_E({\bf r}, {\bf r}')$ is Green's function of
the system in the absence of the exchange coupling. Note that
this function is exact with respect to the potential coupling
$U$ which is not presumed to be weak. The problem of finding
${\cal G}_E$  in the presence of two impurities was solved in
Ref.~\onlinecite{LTM}. The following identity expresses it  in
terms of free electron ($U=0$) Green's function $G_{E}({\bf r},
{\bf r}')$:
\begin{equation}
\label{cal_G_o}
{\cal G}_E^{0}({\bf R},0) = G_E({\bf R},0)\frac{1+2T_EG_E(0)+T^{2}_EG_E^{2}(0)}{1-T^{2}_EG_E^{2}
(0,{\bf R})G_E({\bf R},0)}.
\end{equation}
The $T$-matrix describes the renormalization of the impurity
strength from multiple scattering events,
\begin{equation}
\label{Tmatrix}
T_{E} = \frac{U}{1-UG_{E}(0)},
\end{equation}
whereas $G_E(0)$ denotes Green's function at coinciding
points,
\begin{equation}
\label{Go}
G_E(0) \equiv G_E(0,0)=-\frac{E A_0}{\pi v^2}
\Bigl[\ln{\left(\frac{t}{|E|}\right)}+\frac{i\pi}{2}\Bigr].
\end{equation}
The  area of a graphene unit cell is denoted with
$A_0=3\sqrt{3}a^2/2$ while the Dirac velocity is $v = ta$. The transposed function
${\cal G}_E(0,{\bf R})$ is given by the same expression as Eq.~(\ref{cal_G_o}), where one
simply replaces $ G_E({\bf R},0) \to G_E(0,{\bf
R})$.

Integrating Eq.~(\ref{partial_Jeff}) with respect to {\it
J}, and substituting the expressions for Green's functions,  we
obtain the effective exchange coupling constant in the
integral form,
\begin{equation} \label{Jeff} J_{\rm eff}= 2J^2
\int\limits_{-\infty}^{\infty} \frac{d\omega}{2\pi \hbar}
\frac{\Pi_{i\omega} ({\bf R})}{[(1-U G_{i\omega}(0))^2-U^2\Pi_{i\omega} ({\bf R})]^2},
\end{equation}
where we introduced the following shorthand for the product of two Green's functions,
\begin{equation}
\Pi_{i\omega} ({\bf R}) = G_{i\omega}(0,{\bf R})
G_{i\omega}({\bf R},0)
\end{equation}
In writing Eq.~(\ref{Jeff}) we utilized that time-ordered
Green's functions do not have singularities in the first and
third quadrants of the complex $E$ plane and rotated the
integration path from the real axis counterclockwise to the
angle $\pi/2$ so that it coincides with the imaginary axis, $E
= i\omega$.

Conveniently, this removes the imaginary part of the logarithm
in Eq.~(\ref{Go}), which now becomes,
\begin{equation}
\label{G(0,0)}
G_{i\omega}(0)=-\frac{i\omega A_0}{\pi v^2} \ln{\left(\frac{t}{|\omega|}\right)},
\end{equation}
To determine the coupling constant $J_{\rm eff}$ given by
Eq.~(\ref{Jeff}) it only remains to ascertain the product of
two Green's functions, $\Pi_{i\omega} ({\bf R})$, whose meaning is rather transparent: it determines the
likelihood of a conduction electron to propagate from one
impurity to another and then back to the first one.

The approximation (\ref{G(0,0)}) is valid provided that the
energies are much smaller than the bandwidth, $|E| \ll t$. If,
in addition, the distance between the impurities is much larger
than the lattice spacing, $R \gg a$, a small-momentum expansion
around the two Dirac points in the Brillouin zone is applicable
to the calculation of Green's function. The corresponding
calculations have been performed previously in
Refs.\onlinecite{SAL,LTM}. The result is very sensitive to the
positions of the two points. If both belong to the same
sublattice, one obtains
\begin{equation}
\label{G_0R_G_R0_AA}
\Pi_{i\omega}^{AA} ({\bf R})= -\frac{\omega^2A_0^2}{\pi^2v^4}K_0^2\left(\frac{|\omega|R}{v}\right)\cos^2\theta_{AA},
\end{equation}
where $K_{0}$ is the Macdonald function of the zeroth order.

When two locations belong to the opposite sublattices, a
different expression is found,
\begin{equation}
\label{G_0R_G_R0_AB}
\Pi_{i\omega}^{AB} ({\bf R})
= \frac{\omega^2A_0^2}{\pi^2v^4}K_1^2\left(\frac{|\omega|R}{v}\right)\sin^2\theta_{AB},
\end{equation}
where $K_{1}$ is the Macdonald
function of the first order.
 A different sign, compared with
Eq.~(\ref{G_0R_G_R0_AA}), is  the result of quantum
interference, which is responsible for the opposite signs of interaction in  AA and AB configurations.

Our general result (\ref{Jeff}) can be illustrated first by
calculating the exchange coupling for the weak $U$ limit, $U
\ll \hbar v |{\bf R}|/a^2$, where the denominator in Eq.~(\ref{Jeff}) can be ignored. As a result, one finds ferromagnetic
coupling for the AA configuration of the two impurities,
\begin{equation}
\label{J_AA_pert}
  J_{\rm eff}^{\! AA}({\bf R})=-\frac{J^2}{16\pi \hbar} \frac{ \! A_0^2}{v R^3} \cos^2\!{\theta_{\! AA}}.
\end{equation}
For the AB configuration the coupling is anti-ferromagnetic (and
stronger by a numerical factor),
\begin{equation}
\label{J_AB_pert}
  J_{\rm eff}^{\! AB}({\bf R})=\frac{3J^2}{16\pi \hbar} \frac{ \! A_0^2}{v R^3} \sin^2\!{\theta_{\! AB}}.
\end{equation}
The $1/R^3$ dependence of the two exchange coupling constants,
as well as the signs, are consistent with the results in
the existing literature, though the additional cosine and sine
factors have been overlooked so far. We note that simply
replacing $J$ with $U$ also yields the correct
expressions\cite{LTM} for the potential part of the impurity
interaction, $W({\bf R})$. This is not surprising, since the
perturbation calculations for the two parts of the interaction
are identical.

\section{Strong $U$ limit}

The perturbative couplings (\ref{J_AA_pert})-(\ref{J_AB_pert})
are rather short-range falling off as the third power of the
distance. However, it turns out that as the potential part $U$
of the electron-impurity interaction increases, so does the
range of the exchange coupling. As evident from the form of the
$T$-matrix, the perturbation theory fails if $U \sim \hbar v
R/a^2$, where the integral in Eq.~(\ref{Jeff}) has to be
calculated differently.

\subsection{AA configuration}

Let us first consider the situation of both adatoms residing on
the same sublattice. Since low energies are
important at large distances, $R\gg a$, we can approximate the Macdonald
function as $K_0(x) \approx -\ln{x}$ and write with the help of Eqs.~(\ref{G_0R_G_R0_AA}) and (\ref{Jeff}),
\begin{align}
\label{firstJ_AA_StrongU}
&J_{\rm eff}^{AA}({\bf R})= -\frac{\pi J^2 v^3 R}{ \hbar U^4 A_0^2} \cos^2\!\theta_{AA}\nonumber\\ & \times
\int\limits_{-\infty}^{\infty} \frac{dx\, x^2\ln^2{x}}{\left[ \left(\rho +ix \ln{\left[\frac{R}{a|x|}\right]}
\right)^2
 +\cos^2\!\theta_{AA}\, x^2\ln^2{x}\right]^2},
\end{align}
where we introduced the dimensionless distance $\rho = R \pi v/UA_{0}$. In the weak coupling limit this distance is large, $\rho \gg 1$.
In contrast, in the strong coupling limit this parameter is small. The
integral in Eq.~(\ref{firstJ_AA_StrongU}) converges on $x \sim
\rho \ll 1$, thus justifying the small-argument approximation
used for the Macdonald function $K_0(x)$.

To calculate the integrals in Eq.~(\ref{firstJ_AA_StrongU}) in
the logarithmic approximation, we notice that the logarithms in
the integrand are both large and slow functions of their
arguments. It is therefore tempting to use the standard
approach to such integrals and approximate the logarithms with
their fixed values taken at the characteristic arguments of the
integrand, $x \sim \rho$. It is easy to see, however, that this
would lead to the vanishing of  the integral, as all the
singularities would be located in the same (lower) half-plane
of the complex $x$. It is thus necessary to calculate carefully
the subleading contribution that stems from the variation of
the logarithms with $x$. This calculation is presented in the
Appendix. Its result is
\begin{multline}
\label{AA strong limit}
J_{\rm eff}^{AA}({\bf R})=  \frac{J^2\pi v^2}{4 \hbar A_0 U^3 c \ln^2\! \beta} \Big[ \frac{4c^3\ln (\alpha/\beta)\ln^3 \!\beta}{(\ln^2\! \alpha - c^2 \ln^2\!\beta)^2} \\ - \frac{2c\ln \alpha \ln \beta}{\ln^2\! \alpha - c^2 \ln^2\! \beta} + \ln\Big(\frac{\ln \alpha + c\ln \beta}{\ln \alpha - c\ln \beta}\Big)\Big],
\end{multline}
where we used the shorthands, $\alpha = UA_0/\pi v a$, $\beta =  UA_0/\pi v R$, and $c=|\cos\theta_{AA}|$.
%\begin{equation}
%\label{secondJ_AA_StrongU}
%J_{\rm eff}^{AA}({\bf R})=  \frac{2J^2\pi v^2}{3 \hbar A_o U^3} \frac{\cos^2\theta_{AA} \ln(\frac{UA_{o}}{\pi vR})}{\ln^3 (\frac{UA_{o}}{\pi va})}.
%\end{equation}
%
The obtained result (\ref{AA strong limit}) simplifies in the limit of $\ln \alpha \gg \ln \beta$, corresponding to strong $U$ and large distances $R$, in which case the coupling constant $J_{\rm eff}^{AA}$ becomes,
\begin{equation}
J_{\rm eff}^{AA}({\bf R})=  \frac{2J^2\pi v^2}{3 \hbar A_0 U^3} \frac{\cos^2\theta_{AA} \ln(\frac{UA_{0}}{ vR})}{\ln^3 (\frac{UA_{0}}{va})}.
\end{equation}
The limit $\ln \alpha \gg \ln \beta$ might be difficult to realize (in which case the more general formula (\ref{AA strong limit}) ought to be used), but it illustrates two features of $J_{\rm eff}^{AA}({\bf R})$. First, the coupling is antiferromagnetic: exchange  interaction between adatoms reverses sign compared with the weak-$U$ limit, Eq.~(\ref{J_AA_pert}), quite reminiscent of the potential part $W$ of the adatom interaction\cite{SAL,LTM}.
Second, the coupling constant $J_{\rm eff}^{AA}({\bf R})$ decays very
weakly, logarithmically only. Needless to say, this behavior
extends only to the  distances $R \sim U a^2/\hbar v$, where the
strong coupling limit crosses over to the weak coupling,
Eq.~(\ref{J_AA_pert}), and where both expressions become of the
same order of magnitude (albeit of different sign).

\begin{figure}
        \centering
    	\includegraphics[scale = 0.34]{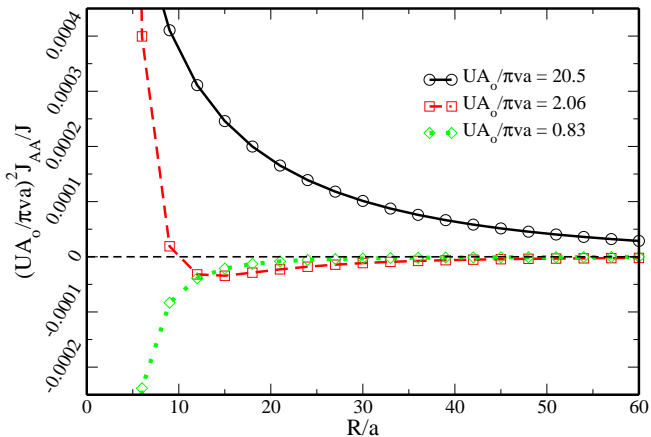}
	\caption{Effective interaction $J_{AA}$ (both impurities belong to the same sublattice) is plotted as a function of distance between the impurities $R/a$ for several value of \textit{U}: 50.0, 5.0 and 2.0 eV (corresponding to the dimensionless ratio $UA_{0}/\pi va$ equal to 20.5, 2.06, 0.83, respectively). $J_{AA}$ is scaled by a factor of coupling constant \textit{J} and dimensionless ratio $UA_{o}/\pi va$. The plot is a result of a numerical integration of the exact effective magnetic interaction \eqref{Jeff}.}
	\label{fig 2}
\end{figure}

Fig.~\ref{fig 2} illustrates the dependence of $J_{\rm eff}^{AA}$ on the distance for different values of the potential coupling strength $U$. For weak $U$ the coupling is ferromagnetic. With increasing  $U$ antiferromagnetic coupling emerges for small distances whereas at large distances ferromagnetic coupling reemerges. With the further increase of $U$ the antiferromagnetic range extends to progressively larger distances.

\subsection{AB configuration}

When the impurities reside on different sublattices, we obtain
from Eqs.~(\ref{G_0R_G_R0_AB}) and (\ref{Jeff}), upon utilizing the approximation for the
Macdonald function, $K_{1}(x) \sim 1/x$,
\begin{align}
\label{firstJ_AB_StrongU}
&J_{\rm eff}^{AB}({\bf R})= \frac{\pi J^2 v^3 R}{ \hbar U^4 A_0^2} \sin^2\!\theta_{AB}\nonumber\\ & \times
\int\limits_{-\infty}^{\infty}  \frac{dx}{\left[ \left(\rho +ix \ln{\left[\frac{R}{a|x|}\right]}
\right)^2
 -\sin^2\!\theta_{AB}\right]^2}.
\end{align}
For small $\rho \ll 1$, the poles of the integrand now reside on the opposite sides of
the real axis. The use of the low-$x$ approximation of the Macdonald function is justified by the fact that
 the integral converges at $x\sim |\sin\theta_{AB}|/\ln(R/a)$. In the leading logarithmic approximation it is sufficient to take
the logarithm at its typical value within the interval of convergence. One thus obtains,
\begin{equation}
\label{secondJ_AB_StrongU}
 J_{\rm eff}^{\! AB}({\bf R})=\frac{J^2}{2\hbar} \frac{ \pi^2 v^3 }{A_0^2 U^4|\sin{\theta_{\! AB}}|} \frac{R}{ \ln(\frac{UA_{0}}{va})}.
\end{equation}
Interestingly, the coupling $ J_{\rm eff}^{\! AB}({\bf R})$ increases with the distance between adatoms. The sign of the coupling is antiferromagnetic, similar to Eq.~(\ref{J_AB_pert}) valid at large distances where the interaction of adatoms is perturbative.

The crossover from strong coupling limit, Eq.~(\ref{secondJ_AB_StrongU}), to the weak coupling limit, Eq.~(\ref{J_AB_pert}) occurs at $\rho \sim 1$. In fact, a resonance takes place at $\rho = \rho_0=|\sin\theta_{AB}|$, which corresponds to a localized impurity level crossing over from one Dirac cone to another, see Eq.~(\ref{energy of a crossing level}). Indeed, the energies of the impurity levels are determined by the zeros of the denominator of the integrand in Eq.~(\ref{Jeff}).

For small values $\rho -\rho_0$ the most important contribution into the integral in (\ref{firstJ_AB_StrongU}) comes from small arguments $x\ll 1$. Keeping only the lowest order terms in $x$ in the denominator of the integrand, we write for the integral in Eq.~(\ref{firstJ_AB_StrongU}), while denoting $\xi = \rho^{2}-\rho_0^{2}$, in the leading logarithmic approximation,
 \begin{align}
\label{strong AB integral}
\int\limits_{-\infty}^{\infty}  \frac{dx}{\left(\xi +2i\rho_0x \ln{(\frac{R}{a|x|}
)} \right)^2}&=-\frac{\partial}{\partial \xi} \int\limits_{0}^{\infty}  \frac{2\xi \, dx}{\xi^2 +4\rho_0^2x^2 \ln^2\!{(\frac{R}{a|x|}
)}}\nonumber\\ &=-\frac{\pi}{2\rho_0|\xi|\ln^2\!{(\frac{R}{a|\xi|}.
)}},
\end{align}
In usual notations Eq.~(\ref{firstJ_AB_StrongU}) now reads,
\begin{equation}
\label{resonantAB}
J_{\rm eff}^{AB}({\bf R})= -\frac{v J^2   }{4 \hbar U^2 |R-R_0|}\frac{|\sin\theta_{AB}|}{\ln^2\!{(\frac{R_0^2}{a(R-R_0)})} }.
\end{equation}

\begin{figure}
        \centering
    	\includegraphics[scale = 0.34]{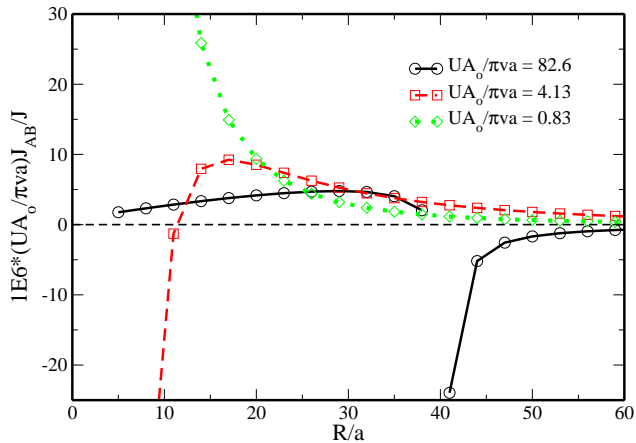}
	\caption{Effective interaction $J_{AB}$ (impurities belong to different sublattices) is plotted as a function of distance between the impurities $R/a$ for several value of \textit{U}: 200, 10 and 2 eV (corresponding to the dimensionless ratio $UA_{0}/\pi va$ equal to 82.6, 4.13, 0.83 respectively). $J_{AB}$ is scaled by a factor of coupling constant \textit{J} and dimensionless ratio $UA_{o}/\pi va$. The plot is a result of a numerical integration of the exact effective magnetic interaction \eqref{Jeff}.}
	\label{fig 3}
\end{figure}

The resonant coupling (\ref{resonantAB}) is ferromagnetic, in contrast to the above considered limits of long distances and short distances, both of which favor antiferromagnetic ordering of impurity spins. The origin of the resonant coupling is the existence of the zero-energy states of the two impurities at distance $R_0$ and the ensuing increase in the scattering of conduction electrons between the impurities.  Fig.~\ref{fig 3} illustrates the exchange coupling in the AB-configuration: the weak antiferromagnetic coupling and the stronger resonant ferromagnetic interaction.

\section{Summary}
The impurities (adatoms) in graphene interact via the exchange of virtual electron-hole excitations. Such interaction has the potential part as well as the effective spin exchange term. The resulting coupling strength $J_{\rm eff}$ is extremely sensitive to the strength of the on-site potential energy $U$ that the conduction electrons experience when they hop on the carbon atom located above the adatom. For weak $U$ the interaction is mediated by the band states and can be treated perturbatively. It is antiferromagnetic, $J_{\rm eff}>0$,  and generally stronger by a numerical factor when two adatoms reside on opposite sublattices, compared with the ferromagnetic coupling, $J_{\rm eff}<0$ for adatoms on the same sublattice. In both instances, when $U$ is weak, the impurity states have large energies and thus play no role when the distance between adatoms significantly exceeds the lattice spacing.

To the contrary, with increasing $U$ the impurity levels move closer to the  Dirac points; as a result, the adatom-adatom interaction is mediated mostly via those levels rather than through the band states of graphene. Even though in the limit of very large $U$ the spin-spin coupling $J_{\rm eff}$ vanishes, its dependence on the distance for finite $U$ becomes highly non-trivial. In the AA-configuration of adatoms, the coupling becomes very long-range decreasing only logarithmically with the distance while being antiferromagnetic in sign -- opposite to the weak coupling limit. In the AB-configuration the presence of the impurity levels results in two surprising features: the appearance of the interval of distances where $J_{\rm eff}$ increases with the distance for $R\ll R_0$ before undergoing a sign reversal and a resonant enhancement with the maximum at $R=R_0$, the distance where one of the impurity levels crosses the Dirac point.

\acknowledgments
We thank Dima Pesin and Oleg Starykh for helpful discussions.
The work was supported by the Department of Energy, Office of Basic
Energy Sciences, Grant No.~DE-FG02-06ER46313.

\appendix*
\section{Calculation of Logarithmic Integrals}
To calculate the integral in Eq.({\ref{firstJ_AA_StrongU}}), let us rescale the integration variable, $x=\rho z$, and introduce the shorthands $\alpha =R/(a\rho)$,  $\beta =1/\rho$,  $c = |\cos\theta_{AA}|$:
\begin{multline}
\label{thirdJ_AA_StrongU}
J_{\rm eff}^{AA}({\bf R})= -\frac{ v^2J^2}{\hbar A_0 U^3}
\int\limits_{-\infty}^{\infty}dz~c^{2}z^{2} \ln^2({\beta}/{|z|}) \\ \times \frac{1}{\Big[\Big\{1+i z\ln{\left(\frac{\alpha}{|z|}\right)}\Big\}^2+c^{2}z^{2} \ln^2(\frac{\beta}{|z|})\Big]^2} \\ =
\frac{J^2 v^{2}c}{2\hbar A_0 U^{3}} \frac{\partial}{\partial c} \int\limits_{-\infty}^{\infty}  \frac{dz}{\Big[\Big\{1+i z\ln{\left(\frac{\alpha}{|z|}\right)}\Big\}^2+c^{2}z^{2} \ln^2(\frac{\beta}{|z|})\Big]}.
\end{multline}
To the logarithmic approximation, we utilize the fact that the integral converges at $z\sim 1$, where $\ln z \ll \ln\alpha, ~\ln\beta$. Expanding the integrand  up to the first order in $\ln z$, we arrive at,
\begin{align}
\label{firstI(A,B)}
J_{\rm eff}^{AA}({\bf R})&= \frac{J^2 v^2 c}{\hbar A_0 U^3} c \frac{\partial}{\partial c} \int\limits_{-\infty}^{\infty} dz \ln |z|\nonumber\\ &\times \frac{iz(1+iz \ln \alpha) + c^{2}z^{2} \ln\beta}{[(1+iz\ln \alpha)^2 + c^{2}z^{2} \ln^2 \beta]^2}.
\end{align}
The last integral has the form $\int_{-\infty}^\infty dz \ln|z| K(z)$, where $K(z)$ is a rational function with all its singularities located in the upper half-plane of complex $z$: this follows from $\alpha > \beta$, and the fact that $c\le 1$. Defining now a new function $Q(z)$ according to $Q(z) =\int_{-\infty}^z dzK(z)$, one can use the integration by parts to obtain,
\begin{align}
\label{integral transformation}
\int\limits_{-\infty}^\infty dz \ln|z| \frac{Q(z)}{dz} &=  -P\int\limits_{-\infty}^\infty dz \frac{Q(z)}{z} \nonumber\\ &=i\pi Q(0) =i\pi \int\limits_{-\infty}^0 dz  K(z).
\end{align}
In performing this transformation we have used that $Q(\infty)=\int_{-\infty}^\infty dzK(z)=0$ since the function $K(z)$ does not have any singularities in the lower half-plane of $z$. Additionally, to express the principal value integral in Eq.~(\ref{integral transformation}) via $Q(0)$, we observe that
\begin{equation}\label{Sokhotsky}
  \int\limits_{-\infty}^\infty dz \frac{Q(z)}{z-i0}= P\int\limits_{-\infty}^\infty dz \frac{Q(z)}{z}+i\pi Q(0) =0,
\end{equation}
as the integral in the left-hand side of Eq.~(\ref{Sokhotsky}) is zero for the already familiar reason: all its poles reside in the upper half-plane.  From Eq.~(\ref{integral transformation}) we obtain that the exchange coupling constant  (\ref{firstI(A,B)}) is expressed in term of the following integral of a rational function,
\begin{equation}
\label{firstI(A,B)1}
J_{\rm eff}^{AA}({\bf R})=  \frac{iJ^2\pi v^2c}{\hbar A_0 U^3} \frac{\partial}{\partial c} \int\limits_{-\infty}^{0} dz \frac{iz(1+iz \ln \alpha) + c^{2}z^{2} \ln \beta}{[(1+iz\ln \alpha)^2 + c^{2}z^{2} \ln^2 \!\beta]^2}.
\end{equation}
The integral is straightforward and ultimately reproduces Eq.~(\ref{AA strong limit}) of the main text.

\end{document}